# Ultrafast terahertz spectroscopy study of Kondo insulating thin film SmB$_6$: evidence for an emergent surface state

Jingdi Zhang[1,2*], Jie Yong[3], Ichiro Takeuchi[4], Richard L. Greene[3], Richard D. Averitt[1,2*]

[1] *Department of Physics, Boston University, Boston, Massachusetts, 02215, USA*

[2] *Department of Physics, University of California San Diego, La Jolla, California, 92122, USA*

[3] *Department of Physics, University of Maryland, College Park, Maryland 20742, USA*

[4] *Department of Material Science & Engineering, University of Maryland, College Park, Maryland 20742, USA*

*Correspondence: jdzhang@physics.ucsd.edu; raveritt@ucsd.edu

**Abstract:**

**We utilize terahertz time domain spectroscopy to investigate thin films of the heavy fermion compound SmB$_6$, a prototype Kondo insulator. Temperature dependent terahertz (THz) conductivity measurements reveal a rapid decrease in the Drude weight and carrier scattering rate at ~T\*=20 K, well below the hybridization gap onset temperature (100 K). Moreover, a low-temperature conductivity plateau (below 20 K) indicates the emergence of a surface state with an effective electron mass of 0.1$m_e$. Conductivity dynamics following optical excitation are also measured and interpreted using Rothwarf-Taylor (R-T) phenomenology, yielding a hybridization gap energy of 17 meV. However, R-T modeling of the conductivity dynamics reveals a deviation from the expected thermally excited quasiparticle density at temperatures below 20 K, indicative of another channel opening up in the low energy electrodynamics. Taken together, these results suggest the onset of a surface state well below the crossover temperature (100K) after long-range coherence of the f-electron Kondo lattice is established.**

In heavy fermion materials, the strong interaction of *5d* conduction band electrons with a Kondo lattice of localized *4f* electrons renormalizes the Fermi surface (FS), resulting in quasiparticles (QP) with enhanced mass. This interaction also opens up a *d-f* hybridization gap at the Fermi surface, resulting in the formation of a narrow insulating gap. Kondo insulators include $SmB_6$, $YbB_{12}$, CeNiSn and spectroscopic techniques including optics have directly observed the hybridization gap [1] [2] [3] [4].

Recent theoretical work has predicted that Kondo insulators are strongly correlated electron systems with the potential to possess a topological non-trivial surface state [5] [6]. The topological surface state emerges due to strong spin-orbit coupling introducing an inversion of the localized 4f band and itinerant 5d bands, a key ingredient for a topological insulator (TI) state [7] [8]. Several experiments on the Kondo insulator $SmB_6$ have investigated the possibility of a non-trivial topological surface state. This includes transport measurements [9], ARPES [10] [11] [12] [13], tunneling spectroscopy [2] [3] [4], and quantum oscillations [14] [15].

Terahertz time-domain spectroscopy (THz-TDS) is a powerful tool to characterize correlated electron materials, providing spectroscopic access to the low-energy electrodynamics [16]. Similarly, ultrafast pump-probe spectroscopy has emerged as an indispensable approach to investigate the response of correlated materials to the ultrafast pulsed electromagnetic excitation that initiates non-equilibrium dynamics [17]. Femtosecond dynamics measurements are particularly sensitive to the opening of gaps or pseudogaps in numerous materials including heavy fermion materials [18] [19] [20] and superconductors [21]. The dynamics can be phenomenologically analyzed using the Rothwarf-Taylor model to gain insight into the quasiparticle density and recovery dynamics [22] [23]. While many of these experiments have used all-optical pump-probe spectroscopy, which probes the low energy electrodynamics indirectly, the combination of THz spectroscopy with sub-picosecond time resolution (i.e. optical-pump THz-probe (OPTP) spectroscopy [16] [24]) is particularly powerful since the temporal response of the low energy electrodynamics is directly measured.

We utilize terahertz time domain spectroscopy to investigate thin films of the heavy fermion Kondo insulator $SmB_6$. Both static THz-TDS measurements and dynamic THz conductivity measurements (using OPTP) are reported. THz-TDS conductivity measurements reveal a rapid decrease in the Drude weight and carrier scattering rate at ~20 K, well below the characteristic temperature (100 K) for opening the hybridization gap. The low-temperature residual conductivity below 20 K indicates the emergence of a surface state with an effective electron mass of $0.1m_e$. Further, the picosecond optical conductivity dynamics exhibit an anomalous amplitude response below 20 K. A Rothwarf-Taylor analysis shows that this behavior arises from a deviation in the thermally excited quasiparticle density expected in a pure single-gap hybridization gap scenario, indicating the opening of an additional channel in the low-temperature electrodynamic response. Our results provide evidence for a crossover from a hybridization gap dominant insulating phase to a surface state dominant regime, suggesting that the onset of a surface state requires long-range coherence of the f-electron Kondo lattice.

In our study, preferentially (001) orientated $SmB_6$ thin films (100 nm thick) were grown by co-sputtering of $SmB_6$ and B targets in ultrahigh vacuum at 800 $^o$C. After in situ annealing of the thin films at the same temperature, the stoichiometry and crystalline orientation was confirmed by wavelength dispersive spectroscopy, TEM and X-ray diffraction [25] [26]. Transport measurements show a resistivity ratio of a factor of two between 2 K and 300 K, due to the emergence of a surface state at low temperatures (**Figure 1 (a)**). **Figure 1(a)** also highlights the different interactions regions in $SmB_6$ as a function of temperature (as determined from point contact spectroscopy [2]), from weakly interacting around 100K to Kondo lattice hybridization below 20K.

A 1 kHz Ti:sapphire regenerative amplifier laser producing 1.55 eV near-infrared pulse (800 nm, 3 mJ, 35 fs) was utilized for these experiments. THz pulses were generated using optical rectification in ZnTe. The THz pulses were used for the temperature dependent THz conductivity measurements and for the OPTP carrier dynamic studies. To measure the THz conductivity of the sample, the temporal waveform (electric field amplitude and phase) of THz pulse was measured with electro-optic sampling [27]. The Fourier transform of THz waveform from sample and reference scans yields the complex conductivity in the frequency domain without Kramers-Kronig analysis. The conductivity dynamics were measured by adjusting the time delay between 1.55 eV excitation pulses and the THz pulses as described elsewhere [24]. **Figure 1 (b)** schematically displays these experiments in terms of the bandstructure resulting from d-f hybridization (blue lines) and surface band (red lines). The far-infrared THz pulse (green) interrogates the low energy electrodynamics (bulk and surface). For the dynamic conductivity measurements, an optical pulse (red) creates an excited carrier distribution that rapidly relaxes to the gap [18-21].

We first consider the far-infrared response in the absence of photoexcitation. **Figure 2 (a)** plots the frequency dependent complex terahertz conductivity ($\sigma_1 + i\sigma_2$) of thin film $SmB_6$ at various temperatures. The real THz conductivity ($\sigma_1$, below ~1 THz) decreases from 1300 $\Omega^{-1}cm^{-1}$ to a saturated value of 800 $\Omega^{-1}cm^{-1}$ from 200 K to 4 K, consistent with DC transport measurements [25]. The observed decrease in $\sigma_1$ is consistent with the opening of the hybridization gap leading to a decrease in the Drude spectral weight. The hybridization gap (HG), however, is at 17 meV (~4.3 THz) and is therefore not spectrally resolved in our measurements (though the dynamics presented below reveal the effects of the HG). The low temperature conductivity ($\sigma_1$ = 800 $\Omega^{-1}cm^{-1}$) corresponds to a 2D sheet conductance of 0.004 $\Omega^{-1}$, consistent with a conductive surface state and sheet resistance given by THz measurement on conventional topological insulators [28] [29].

**Figure 2(b)** presents a zoomed in view of the frequency dependent imaginary conductivity $\sigma_2$ at temperatures from 4 K to 50 K. The linear Drude $\sigma_2$ curves clearly fall into two distinct groupings. In the range from 50 to 25 K the data exhibit similar slopes. A sudden increase in the slope below ~ 20 K is observed, accompanied by a sudden rise in $\sigma_2$ (110 to 140 $\Omega^{-1}cm^{-1}$ at 0.8 THz [30], see inset of **Figure 2(b)**). This behavior in $\sigma_2$ was not observed in previous optical conductivity measurement in single crystal samples [1][31], presumably because the response was dominated by the bulk response. The sudden increase in the slope of $\sigma_2$ is suggestive of a collapse of the carrier scattering rate $\gamma$. In the linear conductivity regime of $\sigma_2$ ($\omega\tau \ll 1$, where $\tau$ is the carrier scattering time) we can roughly estimate the carrier scattering rate $\gamma$ by utilizing the equation: $\gamma = \frac{\sigma_1}{\sigma_2}\omega$.

However, a more precise value of γ can be obtained from Drude model fits of the real and imaginary conductivity. In particular, the experimental THz conductivity at T > 25 K ($\sigma_1$ and $\sigma_2$) is in agreement with a single component Drude oscillator (dashed lines in **Figure 2(a)**). However, below 25 K, $\sigma_1$ starts to deviate from the Drude model at higher frequencies (>1 THz). Nonetheless, a single component Drude fit to the low frequency data (< 1 THz) could be used to extract parameters of the free carriers, including the Drude weight (**Figure 2(c)**) and scattering rate (**Figure 2(d)**), even at temperatures below 25K. It is clear from **Fig. 2(c)** and **(d)** that both the 2D Drude weight ~ $v_p^2 t$ ($v_p$ is the plasma frequency defined as, $v_p^2 t = \frac{ne^2}{4\pi^2 m^*}$, and $t$ is film thickness) and γ show a marked decrease with temperature. The Drude weight decreases ~linearly between 100 K and 30 K then decreases by more than 50% between 30 K to 4 K, saturating at 0.057 (THz²cm). The decrease of $v_p^2 t$ below 30 K could be due to a decrease in carrier density or an increase in effective electron mass. This decrease in $v_p^2 t$ coincides with the sharp decrease of the carrier scattering rate from 8 THz to 4 THz between 20 and 30K (**Fig.2(d)**). A more complete numerical fit that includes an oscillator centered at higher frequency (~ 20 meV or 5 THz) will require an increase of $\sigma_2$ and its slope in the Drude component to compensate the negative contribution due to the high frequency oscillator. Therefore, a single component Drude fit helps to determine the lower limit for the change in scattering rate and Drude weight, suggesting a radical change in the properties of carriers due to the emergence of the surface state. We now turn to conductivity dynamics measurements, which also reveal anomalous non-Kondo-like behavior below 30K.

**Figure 3(a)** displays 1.55 eV photo-excitation induced ultrafast THz conductivity dynamics (Δσ) (or transmission dynamics ΔE/E) as a function of time over the temperature range from 60 K to 5

K. The pump fluence was kept below 9 μJ/cm$^2$ to minimize accumulative thermal heating of the film, confirmed by checking the linearity of the pump-probe dynamics at different fluences. There is an initial increase in the THz conductivity due to photoexcitation of carriers into the conduction band. This is followed by a single exponential relaxation on a picosecond timescale with a flat plateau at longer times. These features are observed in the dynamics at all temperatures investigated. In **Figure 3 (b)**, the data is normalized to the peak, clearly revealing an increase in the photo-excited carrier lifetime as the temperature is decreased. The black lines are fits to the data ($-\Delta E/E = A(T)exp(-t/\tau(T)) + B(T)$). **Figure 3 (c), (d)** plot the lifetime $\tau(T)$ and amplitude $A(T)$ of the fast single-exponential decay dynamics as determined from the fits. The decay time constant increases and then saturates at 10 ps with decreasing temperature, consistent with opening of the hybridization gap at low temperatures. As for the decay amplitude, upon cooling it first displays a monotonic increase for temperatures above 20K. However, an anomaly occurs at lower temperatures (T< 20 K). Instead of reaching monotonically to a saturated amplitude $A(0)$, it decreases by 20% from 20 K to 5 K, in strong contrast to the expected behavior of a single gapped system [23].

The slow down of the single-exponential decay with decreasing temperature arises, in part, from a phonon-bottleneck, characteristic of materials in which an energy gap opens, e.g. superconductors [32] [33], charge density wave materials, [34] and Kondo insulators [19]. As the photoexcited quasiparticles relax, collisional processes lead to phonon generation and excitation of electrons across the hybridization gap Δ. The tendency is for electron-hole recombination to decrease the population of photoexcited carriers. However, the nonequilibrium distribution of phonons with energies in excess of Δ high frequency phonons – HFP) results in electron-hole generation that competes with recombination. This is the origin of the phonon-bottleneck. The plateau in the dynamics (**Fig. 3(a)** and **(b)**) arises from this bottleneck with relaxation of the HFP population on a nanosecond timescale governed by anharmonic decay and thermal transport into the substrate. An additional aspect of importance in the relaxation dynamics is the thermally excited carrier density $n_T$ which decreases with temperature in a gapped system. For low-density excitation (as is the case for the present experiments), the thermally excited carriers play an important role in the relaxation, increasing the availability of electrons and holes that can recombine with photogenerated electrons and holes. That is, with decreasing $n_T$ a longer relaxation time is expected as is observed in our SmB$_6$ films.

The phenomenological *Rothwarf-Taylor* (R-T) model [22] [23] [33] provides a quantitative model to analyze the relaxation dynamics of the photoexcited QPs as heuristically described in the previous paragraph. Specifically, the R-T equations are two coupled differential equations describing the population dynamics of the QPs and HFPs. In the following, we use the R-T model to obtain insight into the underlying physics of the amplitude anomaly. From the temperature dependent single-exponential decay amplitude *A(T)*, we can extract the temperature dependence of the thermally excited carrier (QP) density using

$$n_T \propto \left[\frac{A(T)}{A(0)}\right]^{-1} - 1 \qquad (1)$$

where *A(T)* is the decay amplitude of the QP dynamics, and *A(0)* is the saturated amplitude at T=0. The value of *A(0) = 0.014* is determined by extrapolating the *A(T)* curve from T > 20 K (see **Fig. 3d** dashed line) to zero temperature, corresponding to the expected amplitude in a single gap picture. Moreover, the R-T model also predicts that the decay rate follows

$$\tau = C[D(n_T + 1)^{-1} + n_T] \qquad (2)$$

where C and D are free parameters, and $n_T$ takes the general form

$$n_T \propto T^{1/2} \exp(-\Delta/2k_B T) \qquad (3)$$

where $\Delta$ is the gap energy.

By fitting the decay amplitude and rate (**Figure 4**) simultaneously with Equation (1)-(3), we obtain a Kondo hybridization gap energy of $\Delta$ = 17 meV ± 2 meV, consistent with tunneling [2] [3], ARPES [10] [11] [12] and optical conductivity measurements [1]. From **Figure 4b** it is obvious that, despite the agreement of the R-T model with the experiment at T > 20 K, the experimental QP density deviates from $n_T$ = 0 at (T = 0 K), as predicted by the R-T model for a single gap scenario. This disagreement with the R-T model cannot be reconciled within the framework of the Kondo hybridization gap, since the thermal excited QP density should be strongly suppressed to zero at low temperatures.

Summarizing the results of the static conductivity measurements and the OPTP dynamics, we observe that both experiments reveal an anomalous response below ~20 K. The static THz conductivity shows a sudden transition of both 2D Drude weight and scattering rate of carriers around 20 K. The 1.55 eV optical pump-THz probe non-equilibrium dynamics are governed by display a pronounced bottleneck behavior consistent with a hybridization gap of 17 meV. More importantly, the thermal quasiparticle density $n_T$ deviates from simple R-T model predictions below 20 K, coinciding with the temperature where a sudden change of the Drude weight and scattering rate is observed. In short, the Kondo insulator $SmB_6$ thin films enters a regime where the hybridization gap is insufficient to describe the electrodynamic response below T* = 20 K.

We suggest that the anomalies observed at T* arise from the activation of the conductive surface state and the coherent Kondo lattice, which has also been observed in point contact spectroscopy (PCS) experiment on single crystal $SmB_6$ [2]. PCS experiments suggest that, although the hybridization gap opens up at temperatures up to 100 K, inter-ion correlation emerges at temperature below ~ 30 K, which is close to the T* we observe. To obtain insight to the surface state, we use the low temperature carrier density of the surface state $\sim 2\times10^{14}$ cm$^{-2}$ reported by quantum oscillation (QO) studies [14] to estimate the electron effective mass. Using the QO carrier density with the THz 2D Drude weight, we obtain an electron effective mass of *m\* = 0.1m$_e$*, consistent with theoretical work [35] reporting a much lighter surface state quasiparticle as a result of Kondo breakdown (KB). This further justifies the residual THz conductivity ($v$<1 THz) at low temperature is due to the surface state of $SmB_6$.

We speculate that Kondo coherence state plays a crucial role in the observed electrodynamic anomalies. The onset temperature of the coherent Kondo state [2] is coincides with T* in the current experiment, hinting that Kondo state coherence is a prerequisite for the emergence of the surface state in $SmB_6$. This scenario suggests that at T* the macroscopic surface state emerges and competes with the bulk. This is consistent with the onset temperature of 2D surface state signature in quantum oscillation data [14]. It remains to reconcile these results with ARPES experiments where the electronic band corresponding to surface state emerges at the temperature $T_K$ when hybridization gap opens up.

To conclude, static THz conductivity experiments and ultrafast conductivity measurements display anomalies below a characteristic temperature $T^* = 20K$ strongly indicating the emergence of a surface state that may require the formation of a coherent Kondo lattice for its existence.

**Acknowledgement:** The authors would like to acknowledge Congjun Wu, Gu-Feng Zhang, Xiangfeng Wang, Tao Wu and Xianhui Chen for valuable discussions. JZ and RDA acknowledge support from DOE - Basic Energy Sciences under Grant No. DE-FG02-09ER46643, under which the THz measurements and data analysis were performed. JY, IT and RLG acknowledge support from ONR N00014-13-1-0635 and NSF DMR 1410665.

**Figure Captions:**

Figure 1 (color online). **(a)** Temperature dependent DC resistivity of $SmB_6$ thin film. Regions of different color indicate regimes that $d$ electrons interacts differently with local $f$ electrons as determined by point contact spectroscopy [2]. **(b)** Schematic of THz spectroscopy of Kondo insulator $SmB_6$. Green arrows indicates the low energy excitations that are probed by THz pulse. The 800 nm near-IR pulse initiates ultrafast non-equilibrium dynamics by creating an excited quasiparticle distribution the subsequently relaxes.

Figure 2 (color online). **(a)** Real and imaginary THz conductivity (experiment, solid lines; Drude fit, dash and dotted lines, for real and imaginary part, respectively) at different temperatures (4 K to 200 K). Triangles label the scattering rate of carriers at each temperature. **(b)** Imaginary part of THz conductivity in the temperature range (4 K to 50K). Inset shows $\sigma_{imag}$ at 0.8 THz in the temperature range from 4 to 100 K. **(c)** Temperature dependence of Drude weight. **(d)** Temperature dependence of carrier scattering rate.

Figure 3 (color online). **(a)** Optical (1.55 eV) pump-THz probe dynamics (experiment, thick colored lines; single-exponential fit, thin black lines) of $SmB_6$ thin films from 5 K to 60 K, showing the photo-induced change in transmission (left y-axis) and THz conductivity (right y-axis). **(b)** Normalized pump-probe traces at various temperatures. **(c), (d)** Temperature dependence of the single-exponential decay constant and amplitude. Solid line in **(d)** indicates the temperature dependent decay amplitude predicated by ideal single-gap Rothwarf-Taylor model.

Figure 4 (color online). Rothwarf-Taylor model fitting (solid curve) to **(a)** the single-exponential decay rate and **(b)** thermal excited carrier density $n_T$.

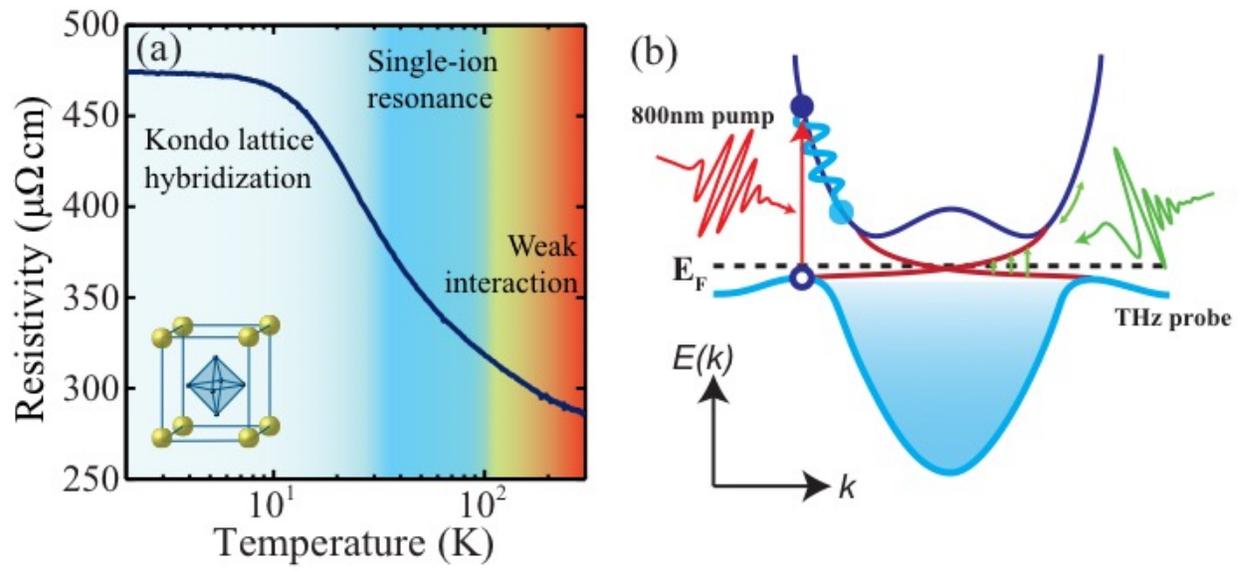

**Figure 1**

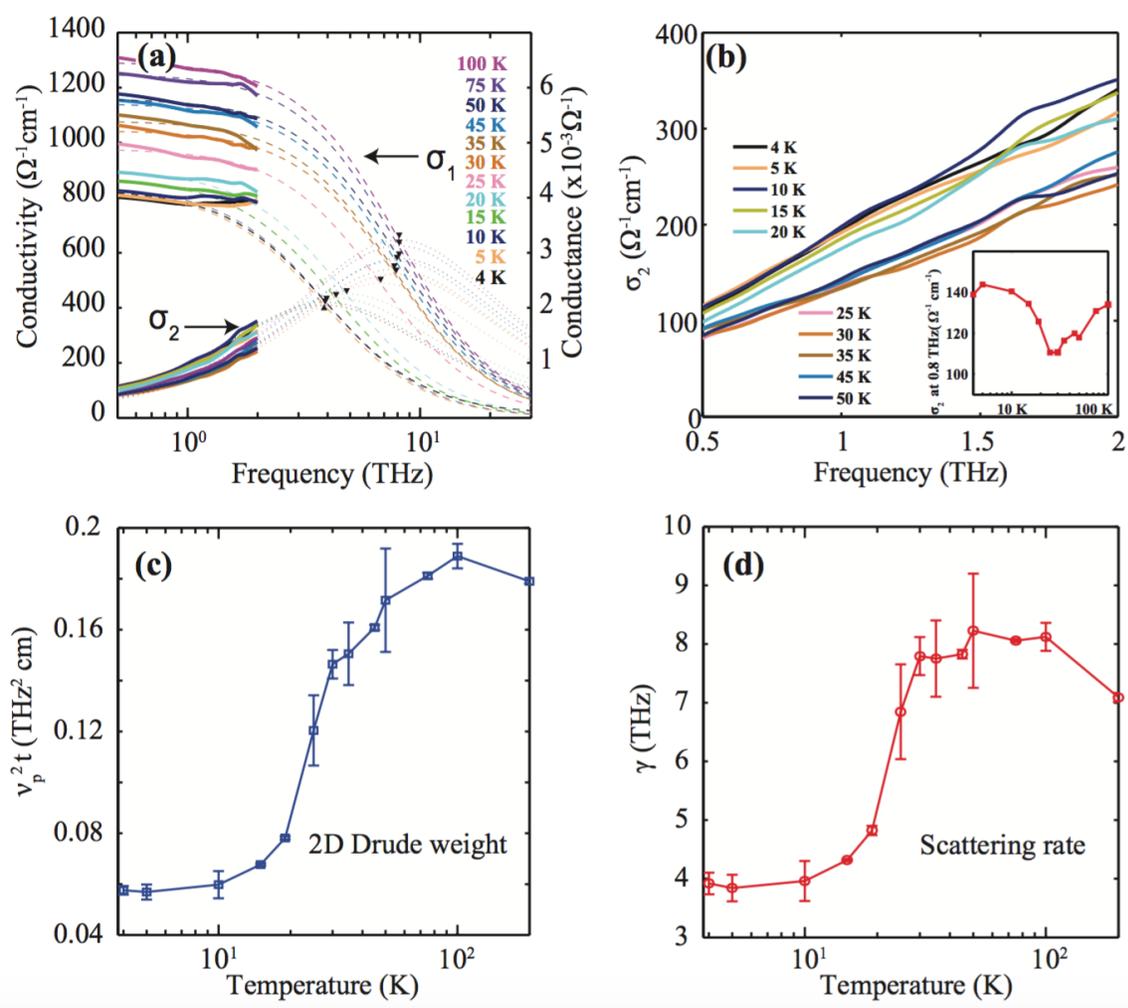

**Figure 2**
16

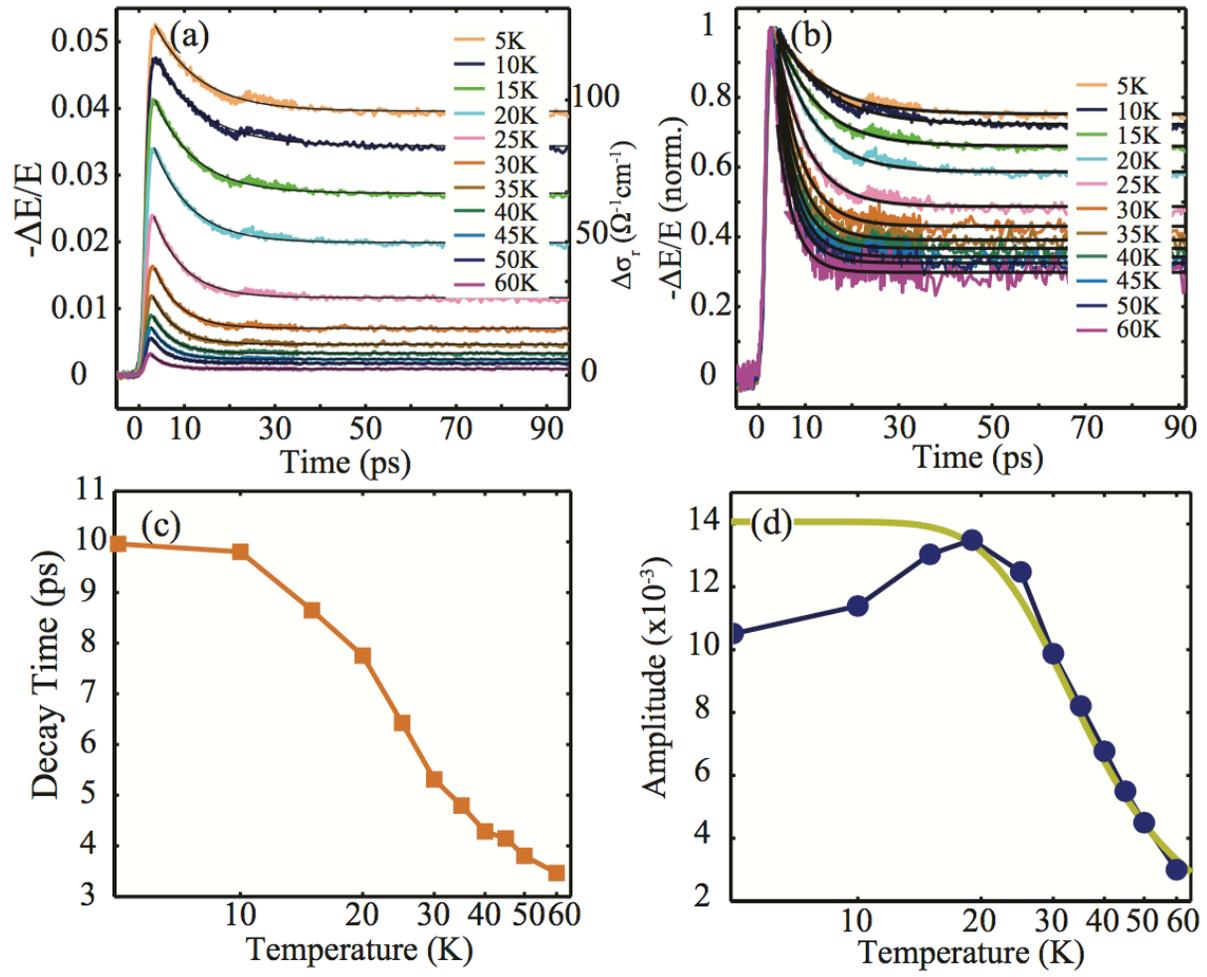

**Figure 3**

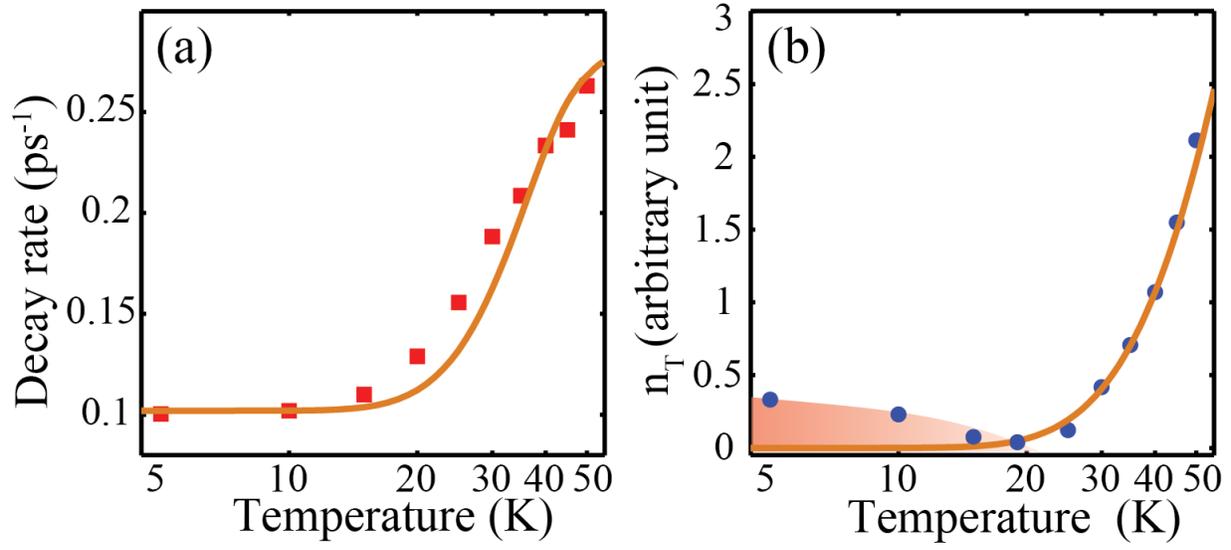

**Figure 4**